\documentclass{JHEP3}
\usepackage{amsmath}
\usepackage{epsfig}
\usepackage{amsmath}

\newcommand{\al} {\alpha}
\newcommand{\p}{{\cal P}}
\newcommand{\N}{{\cal N}}

\newcommand{\ka}{{\kappa}}

\newcommand{\LL}{{\cal L}}

\newcommand{\mat}[1] {\begin{pmatrix}#1\end{pmatrix}}
\newcommand{\bra}[1] {\left<#1\right|}
\newcommand{\ket}[1] {\left|#1\right>}

\title{Schnabl's $\LL_0$ Operator in the Continuous Basis}

\author{Ehud Fuchs and Michael Kroyter\\
Max-Planck-Institut f\"ur Gravitationsphysik\\
Albert-Einstein-Institut\\
14476 Golm, Germany\\
\email{udif@aei.mpg.de, mikroyt@aei.mpg.de}
}

\abstract{ 
Following Schnabl's analytic solution to string field theory, 
we calculate the operators $\LL_0,\LL_0^\dag$
for a scalar field in the continuous $\ka$ basis.
We find an explicit and simple expression for them
that further simplifies for their
sum, which is block diagonal in this basis.
We generalize this result for the bosonized ghost sector,
verify their commutation relation and relate our expressions
to wedge state representations.
} 

\preprint{{\tt hep-th/0605254}\\AEI-2006-034}

\begin{document}

\section{Introduction and summary}

Recently Schnabl found an analytic non trivial solution~\cite{Schnabl:2005gv}
to open bosonic cubic string field theory~\cite{Witten:1986cc}.
This solution is universal~\cite{Sen:1999xm}, and therefore
it was written in a universal way, independent of the CFT of the
matter sector. The only requirement from the matter sector is that it should
have central charge 26 to cancel the central charge of the $b, c$ ghost
system.

Yet, many applications and generalizations of the solution do depend
on the matter sector. Foremost, the proof of Sen's first
conjecture, according to which the height of the potential equals the tension
of a D25-brane, obviously assumes that the matter sector consists of
26 scalars with Neumann boundary conditions.
Yet, for this calculation, the dependence on the matter sector simply amounts
to integrating over the zero-modes, which gives the needed 26 dimensional
volume factor.
Therefore, Schnabl was able to prove Sen's first conjecture within
the universal basis
(up to some subtleties that were clarified
in~\cite{Okawa:2006vm,Fuchs:2006hw}).

Other generalizations, like finding lump
solutions~\cite{Harvey:2000tv,deMelloKoch:2000ie,Moeller:2000jy}
or studying the
close string spectrum around the solution, should also depend on the
matter sector.
Schnabl's solution may also have relevance to other, background
dependent constructions, such as the evaluation of (off-shell)
string amplitudes~\cite{Taylor:2002bq,Taylor:2004rh}.
All this calls for a study of the scalar field in Schnabl's
formulation.

To solve the equations of motion of string field theory, Schnabl made
a specific choice of gauge and coordinates.
The zeroth order Virasoro generator in these coordinates is
\begin{equation}
\label{L0}
\LL_0 = \tan\circ L_0 =\oint \frac{d\tilde z}{2\pi i}\tilde z
    T_{\tilde z\tilde z}(\tilde z) =
    L_0 - \sum_{n=1}^{\infty}\frac{2(-1)^n}{4n^2-1}L_{2n}\,.
\end{equation}
This operator and its conjugate play a prominent role in the solution.
The simplest way to write this operator for a scalar field would
be to use oscillators in the same coordinate system,
\begin{align}
\LL_0 &= \frac{\tilde\alpha_0^2}{2}
  + \sum_{n=1}^\infty \tilde\alpha_{-n}\tilde\alpha_n\,,\qquad
\tilde\alpha_n = \tan\circ \alpha_n =\oint \frac{d\tilde z}{2\pi i}\tilde z^n
    \partial X(\tilde z)\,.
\end{align}
The downside of this approach is that the BPZ conjugate operations become
complicated. For example,
the relation between $\tilde\alpha_n^\dagger$ and $\tilde\alpha_n$
is no longer simple.
Also, the vacuum state annihilated to the right
by all $\tilde\alpha_{n>0}$ remains the same
as can be seen from the relation $U_{\tan}\ket{0}=\ket{0}$.
The conjugate state, the one annihilated to the left by all $\tilde\alpha_{n<0}$,
on the other hand, is the sliver state $\bra{0}U_{\tan}^{-1}=\bra{S}$.
This can also be seen from the generalization of the squeezed state
expression of~\cite{LeClair:1989sp} for the new operators,
\begin{equation}
\bra{f}=\bra{0}\exp\left(\frac{1}{2}\tilde\alpha_n s_{nm} \tilde\alpha_m\right),
  \qquad
s_{nm}=\frac{1}{nm}\oint\frac{dz}{2\pi i}\frac{dw}{2\pi i} z^{-n}w^{-m}
  \frac{f'(z)f'(w)}{\sin(f(z)-f(w))^2}\,.
\end{equation}
Here the trivial map $f(z)=z$ gives the sliver.

Another way to avoid using an explicit expression for $\LL_0$ is to switch
the solution to the Siegel gauge.
Since $L_0=U_{\tan}^{-1}\LL_0 U_{\tan}$, one could naively define,
\begin{equation}
L_0^\ddag \equiv U_{\tan}^{-1}\LL_0^\dagger U_{\tan}\,.
\end{equation}
Then, this operator will satisfy the desired algebra
$[L_0,L_0^\ddag]=L_0+L_0^\ddag$ by construction and one could use these
operators to build a solution equivalent to Schnabl's solution in
the Siegel gauge.
However, trying to calculate $L_0^\ddag$
leads to diverging results, and we did not find a way to make sense
out of this operator.

An alternative route for simplifying Schnabl's operators
stems from their relations to wedge
states~\cite{Rastelli:2000iu,Furuuchi:2001df,Rastelli:2001vb,Schnabl:2002gg}.
That Schnabl's solution is related to wedge states is clear, both
because they are explicitly used in the construction of the solution
and because his gauge choice is implemented via a conformal transformation
which is the inverse of the sliver transformation.
Wedge states are especially easy to deal with in the continuous
basis~\cite{Rastelli:2001hh},
as they are squeezed states whose defining matrix
is diagonal in this basis. In fact, they are the only surface states
with this property~\cite{Uhlmann:2004mv,Fuchs:2004xj}.

The wedge states and the conformal transformation are both generated by
the operators $\LL_0,\LL_0^\dagger$. Thus, one would expect that
these operators would have a nice form in the continuous basis.
To be more concrete, we recall that the wedge state $\ket{n}$
can be represented as,
\begin{align}
\label{Lwedge}
\ket{n}=e^{\log\left(\frac{2}{n}\right){\LL_0^\dagger}}\ket{0}\,,
\end{align}
and also as,
\begin{align}
\label{LLdWedge}
\ket{n}=e^{\big(-\frac{n-2}{2}(\LL_0+\LL_0^\dagger) \big)}\ket{0}.
\end{align}
On the other hand, this should also be equal
to~\cite{Furuuchi:2001df,Rastelli:2001hh},
\begin{align}
\label{OsWedge}
\ket{n}=e^{\int_0^\infty d\ka\, T_n(\ka) a^\dagger_\ka a^\dagger_{-\ka}}
 \ket{0},
\end{align}
where
\begin{align}
\label{Tn}
T_n(\ka)=\frac{e^{\frac{\ka \pi}{2}(n-1)}-e^{\frac{\ka \pi}{2}}}
            {1-e^{\frac{\ka \pi}{2}n}}=
-\frac{\sinh \left(\frac{\ka \pi}{4} (n-2)\right)}
      {\sinh \left(\frac{\ka \pi n}{4}\right)}\,.
\end{align}
Thus, we expect a simple representation for these operators in
the continuous basis.

Yet another hint for the natural description of these operators
in the continuous basis comes from
the commutation relation
\begin{equation}
[\LL_0+\LL_0^\dagger,K_1]=0\,.
\end{equation}
This relation
implies that the bi-linear term of $\LL_0+\LL_0^\dag$ is diagonal
and suggests that the quadratic terms are simple.

Indeed we find,
\begin{equation}
\label{LLdSummary}
\LL_0+\LL_0^\dagger = \frac{\pi}{2}\int_{-\infty}^\infty \frac{d\ka}{\N(\ka)}
  \left(
  2\cosh\big(\frac{\ka\pi}{2}\big)a_\ka^\dagger a_\ka +
  a_\ka^\dagger a_{-\ka}^\dagger +a_\ka a_{-\ka}
  \right).
\end{equation}
Actually, it is very reassuring that we get such a simple and well behaved
result considering the singular behaviour of the Virasoro operators in the
continuous basis~\cite{Douglas:2002jm,Fuchs:2002wk,Belov:2002te}.
It seems that all the singularities conspire to cancel for this
specific combination.
The $\LL_0$ operator itself is slightly less simple
in the sense that in addition to the $\mbox{$\delta$-function}$
contributions it also has $\delta'$ contributions,
\begin{equation}
\label{LSummary}
\LL_0 = \int_{-\infty}^\infty d\ka d\ka'
  \Bigl(\Big(
    \frac{\ka\pi}{4}\coth\big(\frac{\ka\pi}{2}\big)\delta(\ka-\ka')+
    \frac{\ka+\ka'}{2}\delta'(\ka-\ka')\Big) a_\ka^\dagger a_{\ka'} +
  \frac{\pi\delta(\ka+\ka')}{2\N(\ka)}a_\ka a_{\ka'}
  \Bigr).
\end{equation}

These results are derived in section~\ref{sec:evalL0}.
We also derive in this section
the non-zero momentum sector and bosonized ghost form of $\LL_0,\LL_0^\dag$.
In section~\ref{sec:LLcom} we verify that our expressions
indeed satisfy the commutation relation,
\begin{align}
\label{LL}
[\LL_0,\LL_0^\dagger]=&\LL_0+\LL_0^\dagger\,.
\end{align}
We find that the commutation relation holds, up to
regularization
subtleties.
When the central charge is not zero there is an additional infinite
constant on the r.h.s of~(\ref{LL})~\cite{Schnabl:2002gg}.
These infinities cancel between the matter and
the ghost sector, but in the oscillator regularization scheme
we are left with a residual finite constant.
This is reminiscent of the description of wedge states
and string vertices in the continuous
basis~\cite{Belov:2002pd,Fuchs:2002wk,Belov:2002sq,Belov:2003qt,Fuchs:2005ej}.
Section~\ref{sec:WedgeStates} is devoted to relating the different
wedge state representations~(\ref{Lwedge},\ref{LLdWedge},\ref{OsWedge}).
We conclude and suggest future directions in section~\ref{sec:conc}.

\section{The $\LL_0,\LL_0^\dagger$ operators in the continuous basis}
\label{sec:evalL0}

The continuous basis is the basis that diagonalizes
$K_1$~\cite{Rastelli:2001hh}.
For a scalar field the creation and annihilation operators transform as
\begin{equation}
\label{vnka}
a^\dagger_n=\int_{-\infty}^\infty d\ka \frac{v_n^{\ka}}{\sqrt{\N(\ka)}}
  \,a^\dagger_\ka\,, \qquad
a_n=\int_{-\infty}^\infty d\ka \frac{v_n^{\ka}}{\sqrt{\N(\ka)}}\, a_\ka\,.
\end{equation}
The transformation matrix $v_n^\ka$ is defined by
the generating function,
\begin{equation}
f_\ka(z)=\frac{1-e^{-\ka \tan^{-1}z}}{\ka}
  \equiv \sum_{n=1}^\infty\frac{v_n^\ka}{\sqrt{n}}z^n\,,
\end{equation}
and the normalization factor turns out to be~\cite{Okuyama:2002yr}
\begin{align}
\N(\ka)&=\frac{2}{\ka}\sinh(\frac{\ka \pi}{2})\,.
\end{align}
The new creation and annihilation operators obey the commutation relation
\begin{equation}
[a_\ka,a_{\ka'}^\dag] = \rho(\ka,\ka')\,,
\end{equation}
where $\rho(\ka,\ka')=\delta(\ka-\ka')$ is the spectral density.
For $\ka=\ka'$ there are also finite contributions which can
be relevant. We therefore use level truncation to regularize the
$\delta(0)$ contribution,
\begin{equation}
\label{rho}
\rho(\ka)\equiv\rho(\ka,\ka)=
  \lim_{\ell\rightarrow\infty}\frac{1}{2\pi}\sum_{n=1}^{\ell/2}\frac{1}{n} +
  \frac{4\log(2)-2\gamma-\Psi(\frac{i\ka}{2})-\Psi(-\frac{i\ka}{2})}{4\pi}\,.
\end{equation}

\subsection{A direct evaluation of $\LL_0,\LL_0^\dagger$}

Virasoro generators cannot be represented in the continuous
basis by functions, or even by usual delta functions~\cite{Douglas:2002jm}.
This difficulty was addressed
in~\cite{Fuchs:2002wk,Belov:2002te}, where it was shown that
the Virasoro generators can be represented by more general distributions,
i.e. delta functions with complex arguments~\cite{Fuchs:2002wk,Erler:2003eq}.
The positive Virasoro modes are given by,
\begin{align}
L_{m}&= \frac{1}{2}\int_{-\infty}^\infty \frac{d\ka\,d\ka'\,a_\ka a_{\ka'}}
        {\sqrt{\N(\ka)\N(\ka')}} h^{\ka,\ka'}_m+
         \int_{-\infty}^\infty \frac{d \ka d \ka' a^\dagger_{\ka} a_{\ka'}}
                  {\sqrt{\N(\ka)\N(\ka')}} g^{\ka,\ka'}_m\,,
\end{align}
where we refer to the two terms as quadratic and bi-linear, respectively.
The coefficients of these terms, $g^{\ka,\ka'}_m,h^{\ka,\ka'}_m$ are,
\begin{align}
\nonumber
g^{\ka,\ka'}_m=&\sinh\big(\frac{\ka' \pi}{2}\big)
\big(
\frac{q_m(\ka_-)}{\sinh(\frac{\ka_- \pi}{2})}+\\
\label{gkk}
&\frac{i^m \delta(\ka_-{-}2i)-(-i)^m\delta(\ka_-{+}2i)}{2i} 
   -m\sin\big(\frac{m\pi}{2}\big)\delta(\ka_-)   \big),\\
\label{hkk}
h^{\ka,\ka'}_m =&q_m(\ka_+)\,,
\end{align}
where,
\begin{align}
\ka_\pm&\equiv \ka' \pm \ka\,,
\end{align}
and the coefficients depend on,
\begin{equation}
\label{qm}
q_m(\ka)=\frac{1}{2\pi i}\oint \frac{e^{-\ka \tan^{-1}(z)}dz}{z^{m-1}(1+z^2)^2}\,.
\end{equation}
The negative Virasoro generators are simply given by conjugation.

We can now calculate $\LL_0$~(\ref{L0}).
First, we note that all the delta function contributions cancel
between the $L_0$ part and the sum. Thus, we are only left with
the evaluation of,
\begin{equation}
\label{Qdef}
Q(\ka)\equiv -2\sum_{n=1}^\infty \frac{(-1)^n}{4n^2-1}q_{2n}(\ka)\,.
\end{equation}
This would give both the quadratic and the bi-linear coefficient,
since they are both simple functions of $Q(\ka_\pm)$.

We first observe that $Q(\ka)$ is singular for $\ka=0$ since,
$(-1)^n q_{2n}(0)=-n$, and so the sum behaves asymptotically as the
harmonic sum. On the other hand, for $\ka \neq 0$, an oscillatory behaviour
is superposed on the linear divergence of the coefficients and so we
would expect that the series would converge.
The above suggests a component in $Q(\ka)$ which is proportional to
$\delta(\ka)$.
This would also fit our intuition, according to which
$\LL_0,\LL_0^\dagger$ should have a simple form in the continuous basis.

We first use the symmetry of~(\ref{qm}) for even integers to replace the
exponent by a $\cosh$ and then
integrate twice by parts to get,
\begin{equation}
q_{2n}=\frac{2n(2n-1)}{2\pi i}\oint \frac{ \ka \cosh
   \left(\ka \tan ^{-1}(z)\right)+2 z \sinh
   \left(\ka \tan ^{-1}(z)\right)}
   {\ka \left(\ka^2+4\right)z^{2 n+1}}dz.
\end{equation}
In order to perform the sum we have to change the integration contour.
We use the results for these integrals
from~\cite{Fuchs:2002zz,Fuchs:2002wk}, and again integrate one of
the summands by parts to get,
\begin{equation}
q_{2n}=\frac{2 (-1)^{n+1} n (4 n^2-1)
 \sinh \left(\frac{\ka \pi }{2}\right)}{\ka (\ka^2+4) \pi }
 \int_{-\infty}^\infty
   \frac{\cos (\ka u)}{\cosh^4 (u)}\tanh ^{2 n-2}(u)du\,.
\end{equation}
Plugging this result into~(\ref{Qdef}) we get,
\begin{equation}
Q(\ka)=\frac{4\sinh\big(\frac{\ka \pi}{2}\big)}
       {\pi \ka \left(\ka^2+4\right)}
  \sum_{n=1}^\infty n  \int_{-\infty}^\infty
   \frac{\cos (\ka u)}{\cosh^4 (u)}\tanh ^{2 n-2}(u)du\,.
\end{equation}
We can now interchange the order of summation and integration and use
\begin{equation}
\sum_{n=1}^\infty n \tanh ^{2 n-2}(u)=\cosh^4 (u)\,,
\end{equation}
to get
\begin{equation}
Q(\ka)=\frac{4\sinh\big(\frac{\ka \pi}{2}\big)}
       {\pi \ka \left(\ka^2+4\right)}
  \int_{-\infty}^\infty \cos (\ka u)du=
    \frac{4\sinh\big(\frac{\ka \pi}{2}\big)}
      {\pi \ka \left(\ka^2+4\right)}2\pi\delta(\ka)=\pi\delta(\ka)\,.
\end{equation}

Substituting this result into the expression for $h$~(\ref{hkk}) we get that the
term in $\LL_0$ which is quadratic in annihilation operators is,
\begin{equation}
\label{quad}
\LL_{0,\text{quad}}=\pi
   \int_0^\infty \frac{a_\ka a_{-\ka}}{\N(\ka)}d\ka\,.
\end{equation}
The term bilinear in creation and annihilation operators is more
problematic. Here we formally find,
\begin{equation}
\LL_{0,\text{bi-lin}}=
 2\int_{-\infty}^\infty \frac{\sinh\big(\frac{\ka' \pi}{2}\big)}
  {\sqrt{\N(\ka)\N(\ka')}}
  \frac{\delta(\ka-\ka')}{\ka'-\ka}a^\dagger_\ka a_{\ka'} d\ka d\ka'\,.
\end{equation}
This expression contains a very singular distribution that does not
make sense unless it multiplies an expression with a zero at $\ka=\ka'$.
We can, however, define it by a principal part prescription. We would
later justify this choice.
Integration by part and some basic manipulations give,
\begin{equation}
\label{bi-lin}
\LL_{0,\text{bi-lin}}= \int_{-\infty}^\infty \Big(
  \frac{\ka \pi}{4}\coth\big(\frac{\ka \pi}{2}\big)\delta(\ka-\ka')
  +\frac{\ka +\ka'}{2}\delta'(\ka-\ka')\Big)
a^\dag_\ka a_{\ka'} d\ka d\ka'\,.
\end{equation}
We see that the form of $\LL_{0,\text{bi-lin}}$ almost
matches our naive expectation, except for the appearance of a
$\delta'(\ka-\ka')$ term.

Finally, we want to write the sum, $\LL_0+\LL_0^\dagger$ explicitly.
Its quadratic parts are just those of $\LL_0^\dagger$
($\LL_0$) for the creation (annihilation) operators~(\ref{quad}).
For the bi-linear part, we have to sum two expressions.
Here, the antisymmetric part cancels and so,
\begin{align}
\label{LplusL}
(\LL_{0}+\LL_{0}^\dagger)_{\text{bi-lin}}=
\int_{-\infty}^\infty  d\ka \frac{\ka \pi}{2}
  \coth\big(\frac{\ka \pi }{2}\big)
a^\dagger_\ka a_\ka\,,
\end{align}
which is diagonal as it should be.

\subsection{An alternative (half-string) evaluation of $\LL_0+\LL_0^\dagger$}
\label{sec:HalfevalL0}

A useful representation of $\LL_0+\LL_0^\dagger$ is given in
eq.~(2.44,2.45) of~\cite{Schnabl:2005gv},
\begin{align}
\label{LplusLisKK}
\LL_{0}+\LL_{0}^\dagger=
  \frac{2}{\pi}(K_1-2K_1^R)=\frac{2}{\pi}(-K_1-2K_1^L)\,,
\end{align}
where,
\begin{align}
K_1=L_1+L_{-1}\,,
\end{align}
is the operator defining the continuous basis and $K_1^{L,R}$ are
its left and right parts in the half-string formulation.
The operator $K_1$ is trivially diagonal in the $\ka$ basis and
the transformation to half-string basis amounts to mixing
$\pm \ka$~\cite{Erler:2003eq,Fuchs:2003wu}.
These facts imply the block diagonal form of $\LL_0+\LL_0^\dagger$.
We now want to evaluate it directly from~(\ref{LplusLisKK}).

The operator $K_1$ is given in the $\ka$ basis by,
\begin{align}
K_1=-\int_{-\infty}^\infty d\ka \ka a^\dagger_\ka a_\ka\, =
  \int_0^\infty d\ka \ka \left(a^\dagger_{-\ka} a_{-\ka}
    - a^\dagger_\ka a_\ka\right).
\end{align}
We follow here the conventions of~\cite{Fuchs:2003wu} and define
the transformation to the continuous half-string basis by the Bogoliubov
transformation,
\begin{align}
\mat{a^l_\ka \\ a^r_\ka}=
 W \mat{a_{-\ka} \\ a_\ka}+U \mat{a^\dagger_{-\ka} \\ a^\dagger_\ka}.
\end{align}
Here $W,U$ can be any pair of matrices built from block diagonal
rank-one projectors as explained in~\cite{Fuchs:2003wu}.
For the sliver,
\begin{align}
W=\frac{1}{\sqrt{1-e^{-\ka\pi}}}\mat{1&0\\0&1},\qquad
U=\frac{e^{-\frac{\ka\pi}{2}}}{\sqrt{1-e^{-\ka\pi}}}\mat{0&1\\1&0}.
\end{align}
Substituting the inverse transformation
($W\rightarrow W,\,U\rightarrow -U$) in the definition of $K_1$, we
get its form in the (sliver) half-string basis,
\begin{align}
\label{K1h}
K_1^h=\int_0^\infty d\ka \ka \left(a^{l\,\dagger}_\ka a^l_\ka
  - a^{r\,\dagger}_{-\ka} a^r_{-\ka}\right).
\end{align}
The simple $\ka$ dependence is unique to the sliver basis.
Other projectors would result in more complicated expressions that would
also contain bi-linear terms.
The decoupling of the left and right modes is of course general to all half
string bases.
Splitting now to left and right part is obvious.
The first term is $K_1^l$ and the second is $K_1^r$.

We can now use eq.~(\ref{LplusLisKK}) to calculate $\LL_0+\LL_0^\dagger$
in the half-string basis, and then transform back to the $\ka$
basis to get the same result~(\ref{LLdSummary}) we got in the previous
subsection.
In this calculation we get an infinite constant from the
normal ordering of the operators. We study this constant later.

\subsection{Non-zero momentum and the bosonized ghost sector}
\label{sec:ghost}

The operators $\LL_0,\LL_0^\dagger$ contain also a term linear
in the momentum that we still did not calculate. This term would not
contribute in the case of uniform tachyon condensation, because
$p=0$ in this case. However, it may be of importance in studying
generalizations of Schnabl's solution in the context of lump
solutions. Moreover, it is also important for describing the
bosonized ghost sector.

The momentum dependent term in the matter sector
is\footnote{We work in the $2\al'=1$ conventions.},
\begin{align}
\label{delMat}
\delta_1 L_n^\text{m}=\sqrt{|n|} a_n p_0\quad
(n\neq 0)\,,\qquad \delta L_0^\text{m}=\frac{1}{2}p_0^2\,,
\end{align}
which upon using~(\ref{vnka}) implies,
\begin{align}
\label{matLp}
\delta_1 \LL_0^{\text{m}}=\frac{1}{2}p_0^2-
2p_0 \int_{-\infty}^\infty d\ka
 \Big(\sum_{n=1}^\infty
  \frac{(-1)^n\sqrt{2n}\, v_{2n}^\ka}{4n^2-1}\Big)
\frac{a_\ka}{\sqrt{\N(\ka)}}\,.
\end{align}
Complex conjugation gives $\delta_1\LL_0^{\dagger\,\text{m}}$.
In the bosonized ghost sector, $p_0$ should be replaced by the
half-integer ghost number $q_0$, giving
$\delta_1 \LL_0^{\text{g}}$. In this case there are also
additional contributions from
the linear-dilaton character of the bosonized ghost,
\begin{align}
\label{delta2L}
\delta_2 L_n^\text{g}=\frac{Q}{2}(n+1)\sqrt{|n|} a_n\quad
(n\neq 0)\,,\qquad \delta_2 L_0^\text{g}=\frac{Q}{2}q_0\,,
\end{align}
where $Q=-3$.
Thus, in addition to $\delta_1 \LL_0^{\text{g}}$,
$\delta_1 \LL_0^{\dagger\,\text{g}}$, which are complex
conjugate to each other, we have in the ghost sector also,
\begin{align}
\delta_2 \LL_0^{\text{g}}=-\frac{3}{2}q_0+
3\int_{-\infty}^\infty d\ka
 \Big(\sum_{n=1}^\infty
  \frac{(-1)^n\sqrt{2n}\, v_{2n}^\ka}{2n-1}\Big)
\frac{a_\ka}{\sqrt{\N(\ka)}}\,,\\
\delta_2 \LL_0^{\dagger\,\text{g}}=-\frac{3}{2}q_0-
3\int_{-\infty}^\infty d\ka
 \Big(\sum_{n=1}^\infty
  \frac{(-1)^n\sqrt{2n}\, v_{2n}^\ka}{2n+1}\Big)
\frac{a^\dagger_\ka}{\sqrt{\N(\ka)}}\,.
\end{align}
These are not complex conjugate to each other.
We see that we have to evaluate the expressions inside the
parentheses in these two equations. Their difference would then
give~(\ref{matLp}). The problem is, of course, that
these expressions do not converge.

In order to deal with this problem, we separate the sums to
\begin{align}
\sum_{n=1}^\infty \frac{(-1)^n\sqrt{2n}\, v_{2n}^\ka}{2n\pm 1}=
\sum_{n=1}^\infty (-1)^n\frac{v_{2n}^\ka}{\sqrt{2n}}\mp
 \sum_{n=1}^\infty \frac{(-1)^n}{2n\pm 1}
          \frac{v_{2n}^\ka}{\sqrt{2n}}\,.
\end{align}
It was argued in~\cite{Fuchs:2003wu} that the
first term in the r.h.s, which is the divergent one, converges
as a distribution to
\begin{align}
\label{uniDiv}
\sum_{n=1}^\infty (-1)^n\frac{v_{2n}^\ka}{\sqrt{2n}}=
\p\frac{1}{\ka}\,.
\end{align}
The evaluation of the two converging sums is straightforward,
either by using directly the generating function, as was done
in~\cite{Fuchs:2003wu}, or by substituting the integral
representation of $v_{2n}^\ka$~\cite{Fuchs:2002wk},
\begin{equation}
v_{2n}^{\ka}=\frac{(-1)^{n}\sqrt{2n}}{2\pi}\N(\ka)
\int_{-\infty}^\infty
 du\frac{\sin(\ka u)\tanh^{2n-1}(u)}{\cosh^{2}(u)}\,.
\end{equation}
Including~(\ref{uniDiv}) we get,
\begin{align}
\sum_{n=1}^\infty
  \frac{(-1)^n\sqrt{2n}\, v_{2n}^\ka}{2n-1}=
 &\frac{\pi}{2}\p \coth \left(\frac{\ka \pi }{2}\right),\\
\sum_{n=1}^\infty
  \frac{(-1)^n\sqrt{2n}\, v_{2n}^\ka}{2n+1}=
 &\frac{\pi}{2}\p \frac{1}{\sinh \left(\frac{\ka \pi }{2}\right)}\,.
\end{align}
Thus the final result is,
\begin{align}
\delta_1 \LL_0^{\text{g}}=
\delta_1 \LL_0^{\text{m}}=\frac{1}{2}p_0^2-
p_0 \frac{\pi}{2} \int_{-\infty}^\infty d\ka
   \tanh\left(\frac{\ka \pi }{4}\right)
 \frac{a_\ka}{\sqrt{\N(\ka)}}\,,
\end{align}
and the additional ghost terms are,
\begin{align}
\delta_2 \LL_0^{\text{g}}=&-\frac{3}{2}q_0+
\frac{3\pi}{2}\int_{-\infty}^\infty d\ka
 \p \coth \left(\frac{\ka \pi }{2}\right)
\frac{a_\ka}{\sqrt{\N(\ka)}}\,,\\
\delta_2 \LL_0^{\dagger\,\text{g}}=&-\frac{3}{2}q_0-
\frac{3\pi}{2}\int_{-\infty}^\infty d\ka
 \p \frac{1}{\sinh \left(\frac{\ka \pi }{2}\right)}
\frac{a^\dagger_\ka}{\sqrt{\N(\ka)}}\,.
\end{align}

\section{Verifying the commutation relation}
\label{sec:LLcom}

In this section, we evaluate the commutation
relation~(\ref{LL}).
This can be considered as a test to our choice
of the principal part prescription in section~\ref{sec:evalL0}.
We start with the zero-momentum case in~\ref{sec:LLcomQuad},
then we evaluate the linear terms and the bosonized
ghost in~\ref{sec:VerGhost}.

\subsection{The quadratic terms}
\label{sec:LLcomQuad}

The commutation relation,
can be written more explicitly as,
\begin{align}
\label{LLbb}
[\LL_{0,\text{bi-lin}},\LL_{0,\text{bi-lin}}^\dagger]+
[\LL_{0,\text{quad}},\LL_{0,\text{quad}}^\dagger]
=&\LL_{0,\text{bi-lin}}+\LL_{0,\text{bi-lin}}^\dagger\,,\\
\label{LLqq}
[\LL_{0,\text{bi-lin}},\LL_{0,\text{quad}}^\dagger]=&
\LL_{0,\text{quad}}^\dagger\,,
\end{align}
and the conjugate of the last equation.
We start with,
\begin{align}
\label{qqDiv}
[\LL_{0,\text{quad}},\LL_{0,\text{quad}}^\dagger]=
\int_0^\infty d\ka d\ka'
 \frac{\frac{\ka \pi}{2}\frac{\ka' \pi}{2}}
  {\sinh \left(\frac{\ka \pi }{2}\right)
    \sinh \left(\frac{\ka' \pi }{2}\right)}
 (a_\ka a_{\ka'}^\dagger+a_{-\ka'}^\dagger a_{-\ka})\delta(\ka-\ka')\,.
\end{align}
We see that normal ordering this expression brings an infinite factor
of the form of an integral over
$\delta(0)$. This is not much of a surprise, since we are not
working here with the full, $c=0$ theory, but rather with a
$c=1$ one-dimensional matter sector.
We deal with this infinite constant later.
Ignoring the constant term, we write
\begin{align}
\label{LLqqCalc}
[\LL_{0,\text{quad}},\LL_{0,\text{quad}}^\dagger]=
\int_{-\infty}^\infty d\ka
 \frac{\big(\frac{\ka \pi}{2}\big)^2}
  {\sinh^2 \left(\frac{\ka \pi }{2}\right)}a^\dagger_\ka a_\ka\,.
\end{align}
Next we calculate,
\begin{align}
\label{LLbbCalc}
\nonumber
[&\LL_{0,\text{bi-lin}},\LL_{0,\text{bi-lin}}^\dagger]=
\int_{-\infty}^\infty d\ka d\ka'd\ka_1 d\ka_1'
\Big(
  \frac{\ka \pi}{4}\coth\big(\frac{\ka \pi}{2}\big)\delta(\ka-\ka')
  +\frac{\ka +\ka'}{2}\delta'(\ka-\ka')\Big)\cdot\\
\nonumber
&\cdot
\Big(
  \frac{\ka_1 \pi}{4}\coth\big(\frac{\ka_1 \pi}{2}\big)\delta(\ka_1-\ka_1')
  +\frac{\ka_1 +\ka_1'}{2}\delta'(\ka_1-\ka_1')\Big)
\big( a_\ka^\dagger a_{\ka_1} \delta(\ka'-\ka_1') -
 a_{\ka_1'}^\dagger a_{\ka'} \delta(\ka-\ka_1)\big )\\&=
\int_{-\infty}^\infty  d\ka \frac{\ka \pi}{4}
  \frac{\sinh(\ka \pi)-\ka \pi}{\sinh^2\left(\frac{\ka \pi }{2}\right)}
a^\dagger_\ka a_\ka\,.
\end{align}
In order to get to the result, we had to exchange the names of the indices
$\ka \leftrightarrow \ka_1'$, $\ka' \leftrightarrow \ka_1$ for the
expression multiplying the last delta function, evaluate the integral
over $\ka_1'$ and use the identity,
\begin{align}
\label{deltaId}
\delta^{(n)}(x)x^k=\left\{
\begin{array}{cc}
\frac{n!(-1)^k}{(n-k)!}\delta^{(n-k)}(x) & \qquad k \leq n \\
0 & \qquad k > n
\end{array} \right .
\,.
\end{align}
It is immediate that~(\ref{LplusL})
is indeed the sum of~(\ref{LLqqCalc}) and~(\ref{LLbbCalc}).

Finally, we have to verify~(\ref{LLqq}),
\begin{align}
\label{bLinQuad}
[\LL_{0,\text{bi-lin}},\LL_{0,\text{quad}}^\dagger]=\pi
\int_{-\infty}^\infty  d\ka d\ka'
\frac{\frac{\ka \pi}{4}\coth\big(\frac{\ka \pi}{2}\big)\delta(\ka-\ka')
  +\frac{\ka +\ka'}{2}\delta'(\ka-\ka')}
{\N(\ka')}
a^\dagger_\ka a_{-\ka'}^\dagger\,.
\end{align}
Now, we use the invariance of the creation operators under
$\ka\leftrightarrow -\ka'$, to write the second summand in
a symmetric form,
\begin{align}
\nonumber
& \frac{\pi}{2}\int_{-\infty}^\infty  d\ka d\ka'
\frac{\ka +\ka'}{\N(\ka')}\delta'(\ka-\ka')
a^\dagger_\ka a_{-\ka'}^\dagger=\\
& \frac{\pi}{4}\int_{-\infty}^\infty  d\ka d\ka'
(\ka +\ka')\delta'(\ka-\ka')
 \Big(\frac{1}{\N(\ka')}-\frac{1}{\N(\ka)}\Big)
a^\dagger_\ka a_{-\ka'}^\dagger
\,.
\end{align}
Expanding this expression in $\ka-\ka'$ and using once
again the identity~(\ref{deltaId}), brings~(\ref{bLinQuad})
exactly into~(\ref{quad}).

\subsection{The linear terms}
\label{sec:VerGhost}

The additional, momentum-dependent terms should
obey some relations in order that the
algebra~(\ref{LL}) would still hold. In the matter
sector\footnote{These relations are also part of the
ghost sector relations, where we should write
$q_0$ instead of $p_0$. We omit the superscripts m,g in the
following expressions.} the relations are,
\begin{align}
&[\delta_1 \LL_0, \LL_0^\dagger]=
 \delta_1 \LL_0-\frac{1}{2} p_0^2\,,\\
&[\delta_1 \LL_0,\delta_1 \LL_0^\dagger]=p_0^2\,.
\end{align}
Since all the expressions involved are regular,
it is clear that the relations would hold and indeed
they do, as can be seen from a straightforward
calculation.
In the ghost sector the coefficient functions are
singular around $\ka=0$ and the integrals are not
well defined. This should have also been expected,
since the ghost contribution should cancel the
infinite constant from the matter sector.

In the ghost sector some new relations emerge.
Most of them are trivial like the relations that involve
only $\delta_1$. These are,
\begin{align}
&[\delta_1 \LL_0,\delta_2 \LL_0^\dagger]+
 [\delta_2 \LL_0,\delta_1 \LL_0^\dagger]=
 -3q_0\,,\\
&[\delta_2 \LL_0,\LL^\dagger_{0,\text{quad}}]+
 [\LL_{0,\text{bi-lin}},\delta_2 \LL_0^\dagger]=
 \delta_2 \LL_0^\dagger+\frac{3}{2}q_0\,,\\
&[\delta_2 \LL_0,\LL^\dagger_{0,\text{bi-lin}}]+
 [\LL_{0,\text{quad}},\delta_2 \LL_0^\dagger]=
 \delta_2 \LL_0+\frac{3}{2}q_0\,.
\end{align}
The only non-trivial relation is the one related to the
normalization, which we examine next,
\begin{align}
&[\delta_2 \LL_0,\delta_2 \LL_0^\dagger]=
-\Big(\frac{3\pi}{2}\Big)^2\int_{-\infty}^\infty
  d\ka\,\p\frac{\coth \left(\frac{\ka \pi }{2}\right)}
    {\sinh \left(\frac{\ka \pi }{2}\right)\N(\ka)}\,.
\end{align}
This expression diverges, since it has a double
pole in the origin. However, in~\cite{Fuchs:2003wu}
a prescription was given for dealing with this type of
divergences in the continuous basis.
According to this prescription, we should interpret
the following integral in level truncation as,
\begin{align}
\int_{-\infty}^\infty
 d\ka\,\p\frac{2}{\ka^2\N(\ka)}\approx
  \sum_{n=1}^{\ell/2}\frac{1}{n}\,,
\end{align}
where $\ell$ is the oscillator level.
Subtracting this expression inside the integral leaves
us with a converging integral,
\begin{align}
-\Big(\frac{3\pi}{2}\Big)^2\int_{-\infty}^\infty
  d\ka\,\left(\frac{\coth \left(\frac{\ka \pi }{2}\right)}
    {\sinh \left(\frac{\ka \pi }{2}\right)}-
     \frac{2}{\pi^2}\frac{2}{\ka^2}
   \right)\frac{1}{\N(\ka)}=
 -\frac{9}{2}\big(2\log (2)-1\big)\,.
\end{align}
Thus we write,
\begin{align}
\label{linContrib}
&[\delta_2 \LL_0,\delta_2 \LL_0^\dagger]\approx
 -\frac{9}{2}\big(\sum_{n=1}^{\ell/2}\frac{1}{n}+
   2\log (2)-1\big)\,.
\end{align}

We now want to add the contribution to the normalization
coming from~(\ref{qqDiv}) of all 27 modes.
Again, we have to regularize this expression, using the
spectral density~(\ref{rho}),
\begin{align}
\label{quadContrib}
27[\LL_{0,\text{quad}},\LL_{0,\text{quad}}^\dagger]\approx
27\int_0^\infty d\ka \rho(\ka)
 \frac{\left(\frac{\ka \pi}{2}\right)^2}
 {\sinh^2\left(\frac{\ka \pi }{2}\right)} =
   \frac{9}{2}\big
    (\sum_{n=1}^{\ell/2}\frac{1}{n}+2\log(2)-\frac{1}{6}\big)\,.
\end{align}
We can now add all the contributions to the central
term in the commutator of the full matter+ghost operators.
Recall that the terms proportional to the momentum (and
ghost number) already canceled. Thus,
\begin{align}
[\LL_{0,\text{full}},\LL_{0,\text{full}}^\dagger]=
 \LL_{0,\text{full}}+\LL_{0,\text{full}}^\dagger+\frac{15}{4}\,.
\end{align}

There is a discrepancy between this result
and~(\ref{LL}) which comes from using different
regularization schemes. Schnabl did the calculation in
the universal basis with $c=0$ Virasoro operators,
while here we use the continuous basis regularization,
which relies on oscillator level truncation.
There are two potential problems here. The first is the
use of $c\neq 0$ Virasoro algebra due to the separation
to matter and ghost parts and the second is the
use of oscillator level-truncation calculations, which
can lead to anomalous results~\cite{Fuchs:2005ej}.
In fact, already the matter-ghost factorization of
$\LL_0,\LL_0^\dagger$ may
lead to problems, as was noticed
by Schnabl already in~\cite{Schnabl:2002gg}. There,
these operators (called there $B,B^\dagger$) were
used to produce the ``unbalanced wedge states'' and it
was demonstrated that the inclusion of these states in
the algebra may lead to normalization inconsistencies.

We evaluated this constant also in the discrete basis.
The contribution of the linear part~(\ref{linContrib})
is reproduced analytically. This also trivially matches the
contribution that we get in a Virasoro-based level truncation,
since oscillators at level $\ell$ emerge only from
$\delta_2 L_\ell$~(\ref{delta2L}).
The contribution of the quadratic part~(\ref{quadContrib})
was evaluated numerically. We found a finite part, which is
half the one of~(\ref{quadContrib}) to a very high precision.
Thus, the direct use of oscillator level-truncation almost
produces the results of the continuous basis. We believe
that the source of the discrepancy here is that the integrand
of~(\ref{qqDiv}) contains a delta-function, which should also
be interpreted as $\rho(\ka)$. This effectively gives a
product of two such factors. However, we do not know how to
treat such a product. What is really needed is a better
regularization $\rho(\ka,\ka')$~(\ref{rho}) that would allow
such manipulations. A better regularization would presumably
result in a vanishing constant.

Note added: In a later paper we found such a regularization that
indeed resolved this issue.

\section{Wedge states}
\label{sec:WedgeStates}

In this section we demonstrate that our expressions for $\LL_0,\LL_0^\dag$
generate the wedge states via both
relation~(\ref{Lwedge}) and~(\ref{LLdWedge}).
This is another verification of the principal part prescription used
in deriving these expressions.
Since we are working with a $c=1$ system, there are infinite normalization
factors, which we ignore. 

\subsection{Relating the $\LL_0^\dagger$ and squeezed state representations}
\label{sec:Wedge}

The operator $\LL_0^\dagger$ contains terms of the form
$a^\dagger a^\dagger$ and $a^\dagger a$. We can use techniques similar
to those of the appendix of~\cite{Kostelecky:2000hz} to write,
\begin{align}
e^{\frac{1}{2}a^\dagger \tilde S a^\dagger+a^\dagger V a}=
e^{\frac{1}{2}a^\dagger S a^\dagger}e^{a^\dagger V a}\,,
\end{align}
where $\tilde S,S,V$ are matrices and get the relation,
\begin{align}
\label{SVS}
S=\frac{e^{2V}-1}{2V}{\tilde S}\,.
\end{align}
This expression is defined as a power series in $V$. Thus, it is well defined
even for a non-invertible $V$ and since the Taylor expansion has a
linear term, it has a unique inverse.

We want to rewrite the equality between~(\ref{Lwedge}) and~(\ref{OsWedge}) as a
first order differential equation in $n$. Since both expression trivially match
for the initial condition $n=2$, this equation is equivalent to the original
equality, when we use the known $n$ dependence in both
expressions~(\ref{Lwedge}),~(\ref{OsWedge}).
To this end we derive both equations with respect to $n$,
\begin{align}
-n\int_0^\infty d\ka \partial_n T_n(\ka) a^\dagger_\ka a^\dagger_{-\ka} \ket{n}=\int_{-\infty}^\infty d\ka d\ka'
 \big(a^\dagger_\ka a^\dagger_{\ka'}\tilde S(\ka,\ka')+
  a^\dagger_\ka a_{\ka'}V(\ka,\ka')\big)\ket{n}.
\end{align}
We can now plug~(\ref{OsWedge})
into the r.h.s of this equation to get an
expression with only creation operators,
\begin{align}
\label{diffEq}
-n\int_0^\infty d\ka \partial_n T_n(\ka) a^\dagger_\ka a^\dagger_{-\ka} \ket{n}=\int_{-\infty}^\infty d\ka d\ka'
 \big(a^\dagger_\ka a^\dagger_{\ka'}\tilde S(\ka,\ka')+
  a^\dagger_\ka a^\dagger_{-\ka'}V(\ka,\ka')T_n(\ka)\big)\ket{n}.
\end{align}
Now, the coefficients of all sets of creation operators should vanish separately.

Instead of just plugging our solution into~(\ref{diffEq}),
we consider an ansatz of the form of the solution that we
found~(\ref{LSummary}) and show that these equations are
the only solution for the ansatz.
Namely we consider,
\begin{align}
\tilde S(\ka,\ka')&=\tilde S(\ka)\delta(\ka-\ka')\,,\\
V(\ka,\ka')&=V_1(\ka)\delta(\ka+\ka')+V_2(\ka+\ka')\delta'(\ka+\ka')\,.
\end{align}
This can be considered as an independent derivation for the expression
of $\LL_0$.

Integrating the $\delta'$ term gives,
\begin{align}
\nonumber
\int_{-\infty}^\infty & d\ka d\ka' V_2(\ka+\ka')\delta'(\ka+\ka')
 T_n(\ka)a^\dagger_\ka a^\dagger_{\ka'}=\\
-\int_0^\infty & d\ka
 \Big(2 a^\dagger_\ka a^\dagger_{-\ka}
 \big(V_2(2\ka)\partial_\ka T_n(\ka)+T_n(\ka)V_2'(2\ka)\big)+
      T_n(\ka)V_2(2\ka)\partial_\ka(a^\dagger_\ka a^\dagger_{-\ka})\Big)=\\
\nonumber
-\int_0^\infty & d\ka a^\dagger_\ka a^\dagger_{-\ka}
 V_2(2\ka)\partial_\ka T_n(\ka)\,,
\end{align}
where in the last equality we had to integrate by part and thus, in
order to ensure that the boundary terms vanish, we had to assume
that $V_2(\ka)$ times the eigenvalues of $a^\dagger_\ka a^\dagger_{-\ka}$
is not too singular as $\ka\rightarrow 0,\infty$ and that $2\leq n$.
We can now write~(\ref{diffEq}) as an equation that should hold for all
$\ka>0$,
\begin{align}
\label{diffEqSimp2}
-n \partial_n T_n(\ka) =\tilde S(\ka)+2V_1(\ka)T_n(\ka)-
 V_2(2\ka)\partial_\ka T_n(\ka)\,.
\end{align}
All that is left to do now is to plug in the explicit expression for
$T_n(\ka)$~(\ref{Tn}) and expand~(\ref{diffEqSimp2}) in a series
with respect to $n$. Equating the first three non-trivial coefficients,
which are functions of $\ka$ gives a unique choice
for $\tilde S(\ka),V_{1,2}(\ka)$ that exactly
matches~(\ref{quad},\ref{bi-lin}). Plugging these expressions
into the full equation~(\ref{diffEqSimp2}), proves that it is indeed
a solution.

\subsection{Relating the $\LL_0+\LL_0^\dagger$ and squeezed state representations}
\label{sec:SumWedge}

The algebra~(\ref{LL}) was used in~\cite{Schnabl:2005gv} in order
to express the wedge states using the sum
$\LL_0+\LL_0^\dagger$~(\ref{LLdWedge}).
This expression seems very similar to~(\ref{Lwedge}). However,
since the sum $\LL_0+\LL_0^\dagger$ is defined using matrices
which are block diagonal in the $\ka$ basis, it is much easier to
directly derive~(\ref{OsWedge}) from~(\ref{LLdWedge}), since here
there is no problem to use the methods of~\cite{Kostelecky:2000hz}
for the evaluation.

We write $\LL_0+\LL_0^\dagger$ in a block diagonal form as,
\begin{align}
\LL_0+\LL_0^\dagger=A
 \left(a^\dagger \mat{0 & 1 \\ 1 & 0}a^\dagger+
  a \mat{0 & 1 \\ 1 & 0}a\right)+
 C \, a^\dagger\mat{1 & 0 \\ 0 & 1}a
\end{align}
where
\begin{align}
\label{AC}
A=\frac{\ka \pi}
 {4\sinh\left(\frac{\ka \pi }{2}\right)}\,,\qquad
C=\frac{\ka \pi}{2}\coth\left(\frac{\ka \pi }{2}\right)\,.
\end{align}
It is clear that we can express,
\begin{align}
&\exp\left(t(\LL_0+\LL_0^\dagger)\right)=\\
\nonumber
&\exp\big(\eta(t)\big)
\exp\left(\al(t) \, a^\dagger \mat{0 & 1 \\ 1 & 0}a^\dagger\right)
\exp\left(\gamma(t)\, a^\dagger\mat{1 & 0 \\ 0 & 1}a\right)
\exp\left(\al(t) \, a \mat{0 & 1 \\ 1 & 0}a\right).
\end{align}
Since all the matrices involved are commutative, we can use the
appendix of~\cite{Kostelecky:2000hz} to immediately write
differential equations for the unknown functions
$\al(t),\gamma(t),\eta(t)$,
\begin{equation}
\begin{aligned}
\dot \al&=A+2C\al+4A\al^2\\
\dot \gamma&=C+4\al A\\
\dot \al&=e^{2\gamma}A\\
\dot \eta&=2\text{Tr}(A\al)\,,
\end{aligned}
\end{equation}
with the initial conditions,
\begin{align}
\al(0)=\gamma(0)=\eta(0)=0\,.
\end{align}
It may seem that we have too many equations, but given the
initial conditions, the first two imply the third.
We also do not calculate the normalization $\eta$, since we check here
only the matter sector.
Substituting the explicit expressions~(\ref{AC}) into the
equation, the solution is immediate,
\begin{align}
\al(t)=\frac{1}{2}\frac{\sinh\left(\frac{\ka \pi }{2}t\right)}
 {\sinh\left(\frac{\ka \pi }{2}(1-t)\right)}
,\qquad
\gamma(t)=\log\left(\frac{\sinh \left(\frac{\ka \pi }{2}\right)}
 {\sinh \left(\frac{\ka \pi }{2}  (1-t)\right)}\right).
\end{align}
Substituting $t=-\frac{n-2}{2}$ we get
\begin{align}
\al(-\frac{n-2}{2})=\frac{T_n}{2}\,,
\end{align}
where $T_n$ is defined in~(\ref{Tn}). Thus,~(\ref{OsWedge}) is
reproduced, as we wanted to prove.

\section{Conclusions} 
\label{sec:conc} 

We demonstrated in this paper that the operators $\LL_0,\LL_0^\dag$ 
have a simple form for scalar fields in the continuous basis. 
The expressions we found could be used in the study of background 
dependent applications related to Schnabl's solution, but not only. 
Schnabl's choice of coordinates and gauge can also simplify 
calculations around the perturbative vacuum, where again our 
expressions could be put to use. 

It was natural to generalize our results to the bosonized ghost sector.
It is usually the case that the bosonized 
ghosts are
easier to handle in string field theory. 
Still, it would be interesting to generalize our results and 
write
$\LL_0^\text{g},\LL_0^{\dagger\,\text{g}}$ using the $b,c$ ghosts,
especially considering that Schnabl's solution
explicitly relies on them. 

Our construction suffers from regularization subtleties.
It seems to us that the source of the
problems is that the oscillator level truncation that is used for
the regularization of the continuous basis is inconsistent.
Therefore, some calculations, such as the evaluation of wedge states
normalizations cannot be trusted.
Thus, we find the construction of a consistent regularization
scheme for the continuous basis highly desirable.

\section*{Acknowledgments}

We would like to thank Dmitri Belov, Yaron Oz, Robertus Potting,
Martin Schnabl and Stefan Theisen for discussions.
The work of M.~K. is supported by a Minerva fellowship.
The work of E.~F. is supported by the German-Israeli foundation for
scientific research.

\newpage

\bibliography{FK}

\end{document}